\documentclass{PoS}
\pdfoutput=1

\title{Large N reduction in deformed Yang-Mills theories}

\ShortTitle{Large N reduction in dYM theories}

\author{\speaker{Helvio Vairinhos}\\
        Centro de F\'{i}sica do Porto, Universidade do Porto, Porto, Portugal\\
        E-mail: \email{hvairinhos@fc.up.pt}}

\abstract{We explore, at the nonperturbative level, the large N equivalence between ordinary SU(N) Yang-Mills theory on $\mathbb{R}^4$ and on $\mathbb{R}^3 \times S^1$ with double-trace deformations. In particular, we compare the values of the $0^{++}$ glueball mass obtained in both sides of the equivalence.}

\FullConference{31st International Symposium on Lattice Field Theory - LATTICE 2013\\
		July 29 - August 3, 2013\\
		Mainz, Germany}

\begin{document}

\section{Introduction}

In their seminal work \cite{ek}, Eguchi and Kawai noted that the physics of Euclidian $d=4$ SU(N) Yang-Mills theory on an infinite-volume lattice may coincide with the physics at zero volume in the large N limit. Such a volume reduction is possible only if the $Z_N$ symmetry of the theory is preserved in the continuum limit. However, the center symmetry in bosonic lattice gauge theories is often spontaneously broken at small physical volumes, thus invalidating large N reduction in such cases \cite{break}.

One way to avoid the spontaneous breaking of the center symmetry is to reduce the theory down to a finite physical volume that is larger than a critical value (below which the center symmetry naturally breaks) \cite{nn}. Alternatively, a clever choice of twisted boundary conditions allows large N reduction down to zero volume \cite{newtek}.
\\

Recently, \"Unsal and Yaffe proposed \cite{uy} a double-trace deformation of SU(N) Yang-Mills theory living on $\mathbb R^3 \times S^1$ that allows full reduction of the volume of the compact direction without spoiling the centre symmetry. The lattice action of this deformed Yang-Mills (dYM) theory is
\begin{equation} \label{action}
S_{\rm dYM} = -2N^2 b \sum_p {\rm Re}{\rm Tr}~U_p + \frac{1}{L_z^3} \sum_{n=1}^{\lfloor N/2 \rfloor} c_n \sum_{\vec x \in \mathbb R^3} |{\rm Tr}~ \Omega_{\vec x}^n|^2
\end{equation}
where $b = \beta/2N^2 = \lambda^{-1}$ is the inverse lattice 't Hooft coupling, $U_p$ are the plaquettes, $\Omega_{\vec x}$ are the Polyakov loops wrapping $S^1$, $c_n$ are the deformation coefficients, and $L_z$ is the size of the compact direction in lattice units.

We test the validity of the large N equivalence between conventional Yang-Mills on $\mathbb R^4$ and dYM on $\mathbb R^3 \times S^1$ with a fully reduced compact direction. We do so by comparing the $0^{++}$ glueball mass estimated numerically from Monte Carlo simulations of their lattice-regularized theories. 

If such an equivalence holds at large N, then the dYM theory Eq.\ref{action} could be used to study the large N limit of pure Yang-Mills theory at a smaller computational cost, due to the full reduction of one direction.\footnote{Deformed Yang-Mills theories with more than one deformed direction are also possible to define, but they are not easier to simulate. The action of dYM theory on $\mathbb R^{d-k} \times S^k$ includes $O(N^k)$ deformation terms, which makes numerical simulations rather expensive for $k > 1$. While it is possible that only a subset of those deformations might be necessary to keep the center symmetry intact, it is also possible that an excessive number of deformation terms may lead to a non-trivial large N limit for $S_{\rm dYM}$ different from the large N limit of conventional Yang-Mills, at strong coupling.} For the same reason, it could also be used to study the large N meson spectrum.

The preliminary results from our simulations of the dYM theory show a large N glueball mass that is different from the value estimated for Yang-Mills theory \cite{teper}, which is likely due to poor statistics. Further numerical studies of dYM will solve this issue.

\section{Numerical simulations}

We simulate the dYM theory of Eq.\ref{action} with gauge groups $N=3,4,5$, which have at most two independent double-trace deformations. 

In order to test the validity of large N reduction we simulate the dYM theory directly on $L^3 \times 1$ hypercubic lattices with double-trace deformations wrapping the short direction. In our simulations we used $L=8,10,12,16,10,24$.

\begin{figure}[t]
    \centering
        \resizebox{100mm}{!}{
        \includegraphics{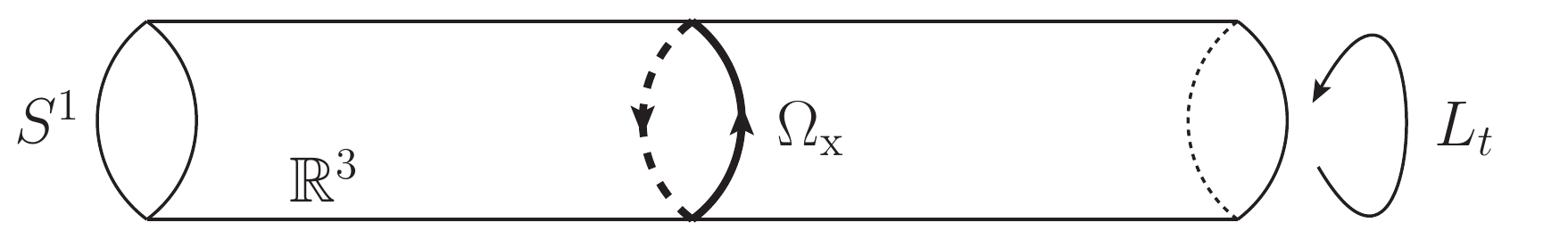}
        }
    \caption{\small\textsf{\textit{Polyakov loop wrapping the compact direction of $\mathbb R^3 \times S^1$.}}\label{fig:R3xS1}
}
\end{figure}

In conventional Yang-Mills, such lattices would be in the deconfined phase, which is characterized by a non-zero expectation value of the Polyakov loop wrapping the smallest direction (Fig.\ref{fig:R3xS1}). In dYM, however, the centre symmetry is intact for sufficiently large values of the deformation coefficients $c_n$, and a physical confining phase is expected to exist for sufficiently small $b$. In our simulations we use the conservative values $c_1 = 2.0$ and $c_2 = 0.5$ \cite{center}.

\subsection{Pseudo-heatbath algorithm}

Since the lattice action Eq.\ref{action} is nonlinear with respect to the link variables (due to the double-trace terms), the Cabibbo-Marinari pseudo-heatbath algorithm \cite{cm} cannot be implemented directly (it requires that the lattice action is linear with respect to each link variable).

However, it is possible to linearize Eq.\ref{action} with respect to the link variables by adding a set of auxiliary Gaussian degrees of freedom to the action and a set of Hubbard-Stratonovich transformations that remove the nonlinear terms \cite{heatbath}. In that situation, the Cabibbo-Marinari pseudo-heatbath algorithm can then be applied directly. We use such an algorithm in our simulations, which is faster and decorrelates quicker than the Metropolis alternative.

Our preliminary calculations involved the generation of $O(10^6)$ thermalized configurations for each lattice spacing and each $N$.

\subsection{$0^{++}$ glueball masses}

We estimate the mass of the $0^{++}$ glueball in lattice units from the two-point function of appropriately smeared and blocked spatial plaquette operators, $\bar u_p$, whose quantum numbers match those of the $0^{++}$ glueball state:
\begin{equation}
  \langle \bar u_p(0)^\dag \bar u_p(\tau) \rangle \propto e^{-a m_{0^{++}}\tau} + \cdots
\end{equation}

To set the scale, we estimate the string tension in lattice units, $a\sqrt\sigma$, from the mass of the torelon states, $a m_l$. The torelon masses are extracted from two-point functions of appropriately smeared and blocked $\vec p=0$ spatial Polyakov loops $\bar l$ wrapping non-compact directions:
\begin{equation}
  \langle \bar l(0)^\dag \bar l(\tau) \rangle \propto e^{-am_l\tau} + \cdots
\end{equation}
The string tension in units of the lattice spacing is then estimated assuming a L\"uscher correction:
\begin{equation}
L a^2 \sigma = a m_l + \frac{\pi}{3L} + \cdots
\end{equation}

In our simulations we choose values of the lattice coupling that correspond to sizes of the non-compact directions larger than 1 fm, so that the system is in the confined phase. 

\begin{figure}[ht]
    \centering
        \resizebox{130mm}{!}{\includegraphics{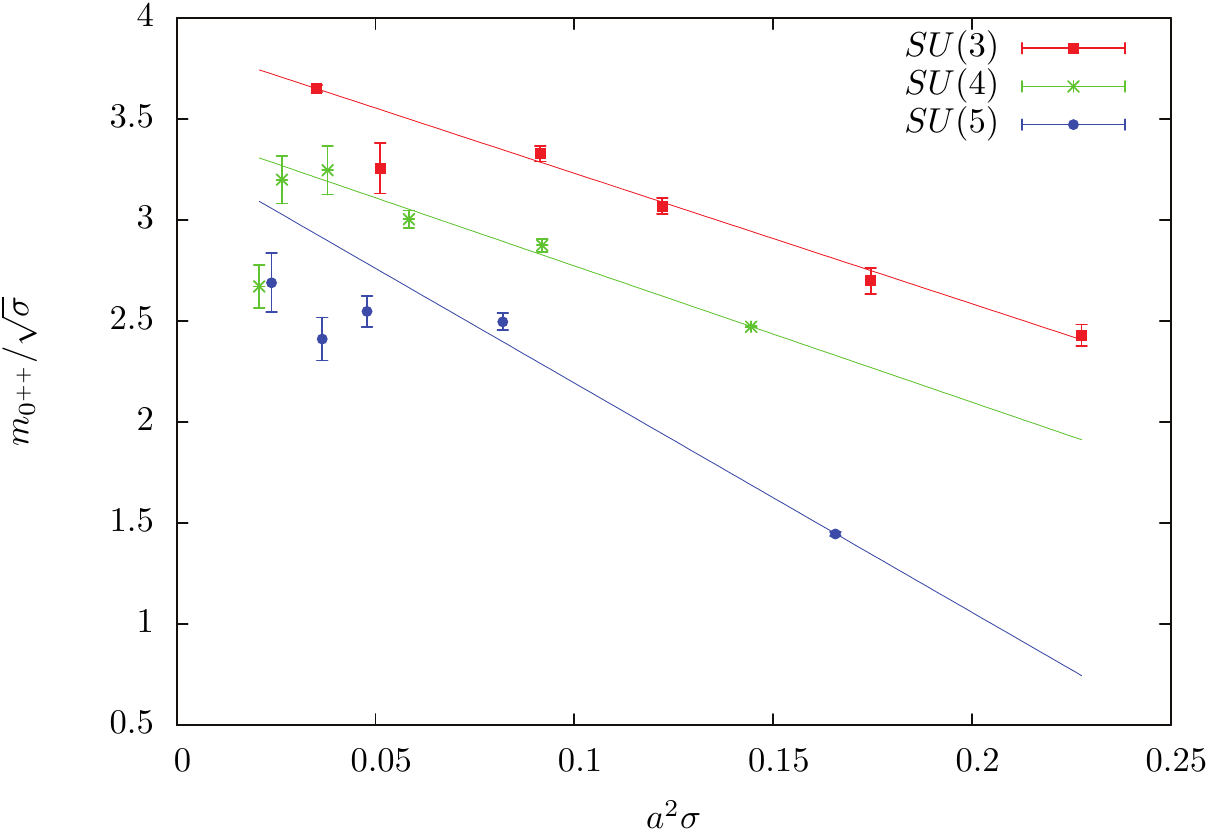}}
    \caption{\small\textsf{\textit{Continuum limit of the $0^{++}$ glueball masses for several gauge groups.}}\label{fig:continuum}
}
\end{figure}

The dimensionless ratios $m_{0^{++}}/\sqrt\sigma$ for various lattice sizes and gauge groups are given in Fig.\ref{fig:continuum}. The continuum limit extrapolations were performed after removing the outliers.

\subsection{Large N limit}

\begin{figure}[ht]
    \centering
        \resizebox{130mm}{!}{\includegraphics{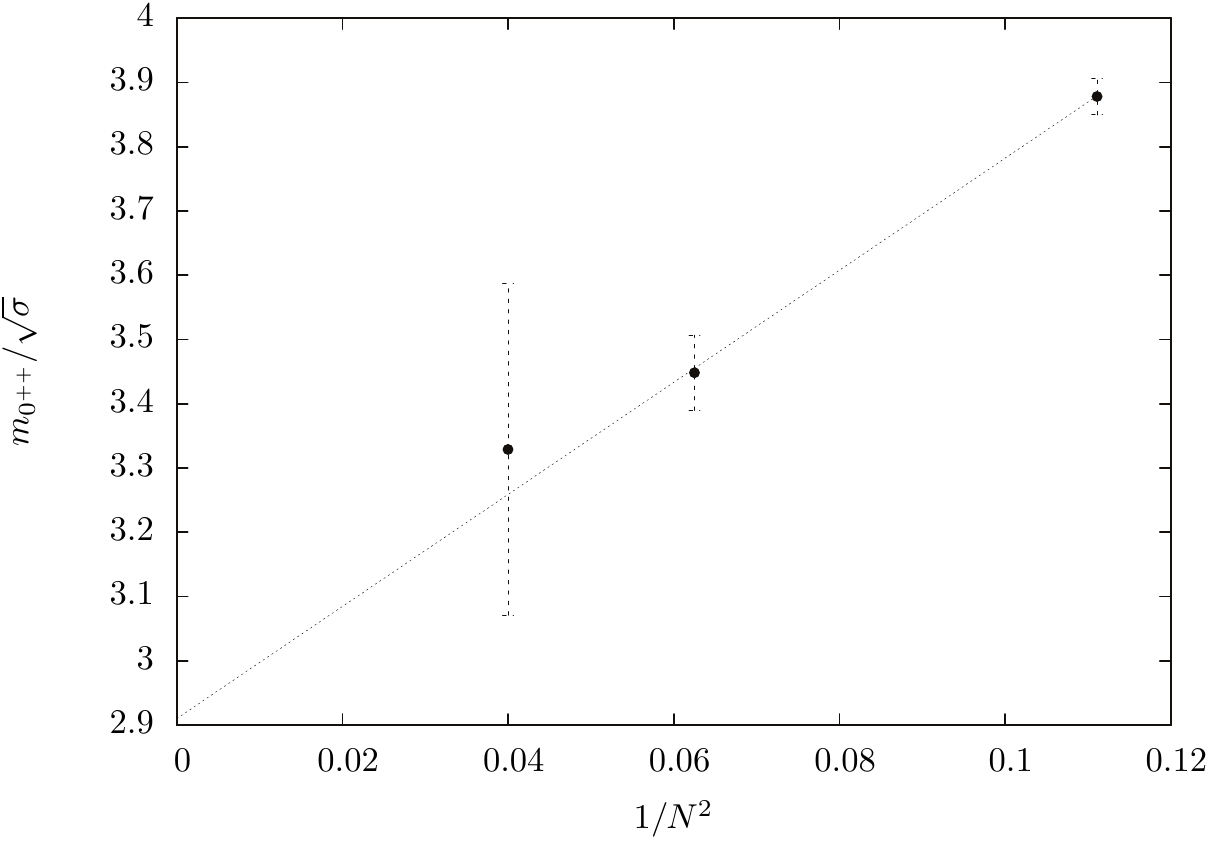}}
    \caption{\small\textsf{\textit{Large N limit of the $0^{++}$ glueball masses.}}\label{fig:largeN}
}
\end{figure}

After extracting the continuum values of the $0^{++}$ glueball masses, we take their large N limit assuming a leading 1/N$^2$ correction (Fig.\ref{fig:largeN}),
\begin{equation}\label{dYM_mass}
  \left.\frac{m_{0^{++}}}{\sqrt\sigma}\right|_N \approx 2.91(4) + \frac{8.72(4)}{N^2}
\end{equation}
which we compare with the result obtained from simulations of Ytheory \cite{teper},
\begin{equation}\label{YM_mass}
  \left.\frac{m_{0^{++}}}{\sqrt\sigma}\right|_N \approx 3.37(15) + \frac{1.93(85)}{N^2}
\end{equation}

There is a discrepancy in both the large N limit of the glueball mass and, more pronouncedly, in the leading-order 1/N correction.

\section{Conclusion}

We observe a discrepancy between the large N values of the $0^{++}$ glueball mass calculated in dYM and in conventional Yang-Mills \cite{teper}. This is possibly due to deficient statistics in our estimations (which produced a number of outlier points that we disregarded in the extrapolations). A more precise study will be presented elsewhere, in order to improve the accuracy of our results, and to determine whether large N equivalence between Yang-Mills and dYM holds.

We observe an even larger discrepancy between the values of the leading 1/N correction in Yang-Mills and dYM. Such a discrepancy is natural, since the theories are very different at finite N, and there is no reason to expect their 1/N corrections to be related. However, such a large value for the leading 1/N correction in dYM means that this theory is not as close to its large N limit than conventional Yang-Mills is. Precise estimations of large N observables from dYM might also require simulations for larger values of N than usual.

We plan to increase the statistics of the present study and analyse the nonperturbative spectrum of dYM in more detail, in order to quantify unambiguously the validity of large N reduction in the presence of double-trace deformations.

\section{Acknowledgements}

We would like to thank to Masanori Hanada, Marco Panero, Mithat \"{U}nsal, Mike Teper and Jo\~{a}o Penedones for stimulating interactions during the development of this work. We would also like to thank the warm hospitality at the Yukawa Institute for Theoretical Physics, where part of this work was developed. Our lattice calculations were carried out on the Avalanche cluster at FEUP/University of Porto. 

HV is supported by FCT (Portugal) under the grant SFRH/BPD/37949/2007 and the project CERN/FP/123599/2011.

\end{document}